\documentclass[10pt,conference]{IEEEtran}
\IEEEoverridecommandlockouts
\usepackage[T1]{fontenc}
\usepackage{cite}
\usepackage{amsmath,amssymb}
\usepackage{graphicx}
\usepackage{url}
\usepackage{enumitem}
\usepackage{balance}
\usepackage[hidelinks,breaklinks=true]{hyperref}
\usepackage{listings}
\usepackage{tikz}
\usepackage{tabularx}
\usetikzlibrary{positioning,arrows.meta,shapes.geometric,fit,backgrounds,calc,shadows}
\lstdefinestyle{artifact}{
  basicstyle=\ttfamily\footnotesize,
  columns=fullflexible,
  keepspaces=true,
  breaklines=true,
  showstringspaces=false,
  frame=single,
  framerule=0.3pt,
  rulecolor=\color{black!25},
  backgroundcolor=\color{black!03},
  captionpos=b,
  aboveskip=4pt,
  belowskip=6pt,
  xleftmargin=0pt,
  xrightmargin=0pt
}

\usepackage{array}
\usepackage{booktabs}

\usepackage[absolute]{textpos}
\usepackage{comment}

\begin{document}





\newcommand{\papernote}{This paper will be presented at IEEE VLSI Test Symposium (VTS) 2026}

\IEEEoverridecommandlockouts
\IEEEpubid{\makebox[\columnwidth]{979-8-3315-6337-0/26/\$31.00 \copyright2026 IEEE \hfill} \hspace{\columnsep}\makebox[\columnwidth]{}}
\begin{textblock}{5}(11.8,0.8)
(Special Session)
\end{textblock}

\title{%
{\normalfont\normalsize \papernote \par}
\vspace{-1mm}
AI-Assisted Hardware Security Verification: A\\
Survey and AI Accelerator Case Study
\vspace{-2mm}
}

\author{
\IEEEauthorblockN{\small Khan Thamid Hasan, Md Ajoad Hasan, Nashmin Alam,
Md. Touhidul Islam, Upoma Das, Farimah Farahmandi}
\IEEEauthorblockA{\textit{\small Department of Electrical and Computer Engineering, University of Florida, Gainesville, FL, USA} \\
\{\small khanthamidhasan, md.hasan, nashminalam, mdtouhidul.islam, upoma.das\}@ufl.edu,
farimah@ece.ufl.edu}
\thanks{We thank the U.S. National Science Foundation (NSF) for support through CAREER Award No. 2339971.}
}

\IEEEaftertitletext{\vspace{-2.5\baselineskip}}

\maketitle

\begin{abstract}

As hardware systems grow in complexity, security verification must keep up with them. Recently, artificial intelligence (AI) and large language models (LLMs) have started to play an important role in automating several stages of the verification workflow by helping engineers analyze designs, reason about potential threats, and generate verification artifacts. This survey synthesizes recent advances in AI-assisted hardware security verification and organizes the literature along key stages of the workflow: asset identification, threat modeling, security test-plan generation, simulation-driven analysis, formal verification, and countermeasure reasoning. To illustrate how these techniques can be applied in practice, we present a case study using the open-source NVIDIA Deep Learning Accelerator (NVDLA), a representative modern hardware design. Throughout this study, we emphasize that while AI/LLM-based automation can significantly accelerate verification tasks, its outputs must remain grounded in simulation evidence, formal reasoning, and benchmark-driven evaluation to ensure trustworthy hardware security assurance.

\end{abstract}


\section{Introduction}
Hardware security verification remains a high-stakes requirement for modern digital systems, as vulnerabilities introduced before fabrication can remain undetected into deployment. The threat landscape covers hardware Trojans, power and timing side channels, unauthorized information flows, and attacks against logic locking and obfuscation, while the verification toolbox still depends on labor-intensive simulation, assertion-based verification, model checking, and leakage assessment methodologies \cite{tehranipoor2010,bhunia2014,kocher1999,mangard2007,hua2018,hu2011,tiwari2009,rajendran2012,subramanyan2015,bushnell2000,ieee1800_2017,accellera_uvm,clarke1999,vijayaraghavan2005,goodwill2011,schneider2015}.

These challenges appear across a wide range of hardware platforms. For instance, AI accelerators embody valuable model parameters, sensitive activations, privileged control paths, and rich microarchitectural side effects, making them attractive targets for side-channel extraction, fault injection, neural Trojan insertion, and accelerator-specific intellectual property (IP) theft \cite{dnnhwsurvey2021,lee2025secureml,batina2019,mercury2023,horvath2025barracuda,powergan2023,gupta2023aiattacksai,rakin2019,derailed2024,10.1145/3583781.3590242,li2024lldnn}. As AI accelerators are integrated into edge, datacenter, and safety-critical platforms, scalable pre-silicon assurance becomes increasingly important.

Recent progress in AI and LLMs has also changed the way hardware security verification can be organized and executed. Early surveys and position papers argue that LLMs can assist security reasoning across the verification stack, but they also warn that trust, grounding, and hallucination remain central concerns \cite{saha2023llm_soc_security,akyash2024evolutionary,abdollahi2024hwverif_llm_survey,wang2024llm_chipdesign_security}. In parallel, LLMs have already shown strong utility in adjacent hardware-design tasks such as register-transfer level (RTL) generation, domain adaptation, self-correcting testbench generation, coverage-oriented prompting, and hallucination mitigation \cite{verigen2024,liu2023rtllm,rtlcoder2024,chipnemo2023,10992873,llm4cov2025,10993072}. Those capabilities naturally motivate their extension into security-oriented planning, property generation, vulnerability analysis, and workflow orchestration.

\begin{figure*}[!t]
\centering
\begin{tikzpicture}[
  >=Stealth,
  node distance=0.55cm,
  stage/.style={
    draw, rounded corners=3pt, minimum height=1.05cm, minimum width=2.45cm,
    text width=2.3cm, align=center, font=\scriptsize\bfseries,
    fill=#1, draw=#1!85!black, text=white,
    drop shadow={shadow xshift=0.4pt, shadow yshift=-0.4pt, opacity=0.18}
  },
  approach/.style={
    draw, rounded corners=3pt, minimum height=1.05cm, minimum width=2.45cm,
    text width=2.3cm, align=center, font=\tiny,
    fill=#1!10, draw=#1!60!black, text=black
  },
  arr/.style={->, thick, color=black!70},
  feedarr/.style={->, densely dashed, thick, color=black!40},
  assistlink/.style={densely dotted, thick, color=black!35},
]

\node[stage=blue!70!black] (s1) {Asset\\Identification};
\node[stage=teal!70!black, right=of s1] (s2) {Threat\\Modeling};
\node[stage=green!50!black, right=of s2] (s3) {Test Plan\\Generation};
\node[stage=purple!70!black, right=of s3] (s4) {Simulation \&\\Formal Verif.};
\node[stage=red!65!black, right=of s4] (s5) {Counter-\\measures};

\draw[arr] (s1) -- (s2);
\draw[arr] (s2) -- (s3);
\draw[arr] (s3) -- (s4);
\draw[arr] (s4) -- (s5);

\node[approach=blue!70!black, below=0.72cm of s1] (l1) {Asset extraction,\\refinement, and\\specification-guided reasoning\\\cite{hasan2026lassetllmassistedsecurityasset,lashed2025,svllm2025}};
\node[approach=teal!70!black, below=0.72cm of s2] (l2) {Threat synthesis,\\CWE-aware reasoning,\\vulnerability analysis, and planning\\\cite{blocklove2023chip,10462177 ,paria2023divas,11050068,11022932,svllm2025,11022958,lashed2025}};
\node[approach=green!50!black, below=0.72cm of s3] (l3) {Threat-to-test translation,\\coverage-guided generation,\\and context-aware planning\\\cite{11022932,svllm2025,zhang2023llm4dv,chipnemo2023,10323953}};
\node[approach=purple!70!black, below=0.72cm of s4] (l4) {Assertion and property synthesis,\\proof assistance, etc. \\\cite{fang2024assertllm,kande2024tifs,maddala2024laagrv,bai2025assertionforge,lyu2025assertgen,wu2025spec2assertion,rajabi2025stellar,wang2025deepassert,lyu2026assertminer,paul2025lisa,pulavarthi2025assertionllm,orenesvera2023rtl2sva,orenesvera2021autosva,qayyum2024incrementalproof,paria2023divas,11050068,ankireddy2025lasa,lasp2024,11022949,11022958,llm4sechw2023,10682659,lashed2025,meic2024,rostami2024chatfuzz}};
\node[approach=red!65!black, below=0.72cm of s5] (l5) {Repair and patching,\\leakage analysis, red-teaming,\\and mitigation co-design\\\cite{10462177 ,pearce2023,gptpresilicon2023,ghost2025,kokolakis2024llmtrojan,svllm2025,11050068,wang2024llm_chipdesign_security}};

\draw[assistlink] (l1) -- (s1);
\draw[assistlink] (l2) -- (s2);
\draw[assistlink] (l3) -- (s3);
\draw[assistlink] (l4) -- (s4);
\draw[assistlink] (l5) -- (s5);

\begin{scope}[on background layer]
  \node[fit=(l1)(l5), inner sep=6pt, draw=black!15, fill=black!3, rounded corners=4pt] (assistband) {};
\end{scope}
\node[below=0.15cm of assistband.south, font=\scriptsize\itshape, text=black!65] {Some LLM-based approaches discussed in this survey};

\node[above=0.10cm of s3, font=\small\bfseries, text=black!80] {Security Verification Pipeline};

\node[left=0.38cm of s1, font=\tiny\itshape, text=black!55, text width=1.45cm, align=right] {RTL, specs,\\and security intent};
\node[right=0.38cm of s5, font=\tiny\itshape, text=black!55, text width=1.55cm, align=left] {Evidence-backed\\claims and mitigations};

\end{tikzpicture}
\vspace{-8mm}
\caption{End-to-end AI/LLM-assisted security verification pipeline. }
\vspace{-6mm}
\label{fig:pipeline}
\end{figure*}
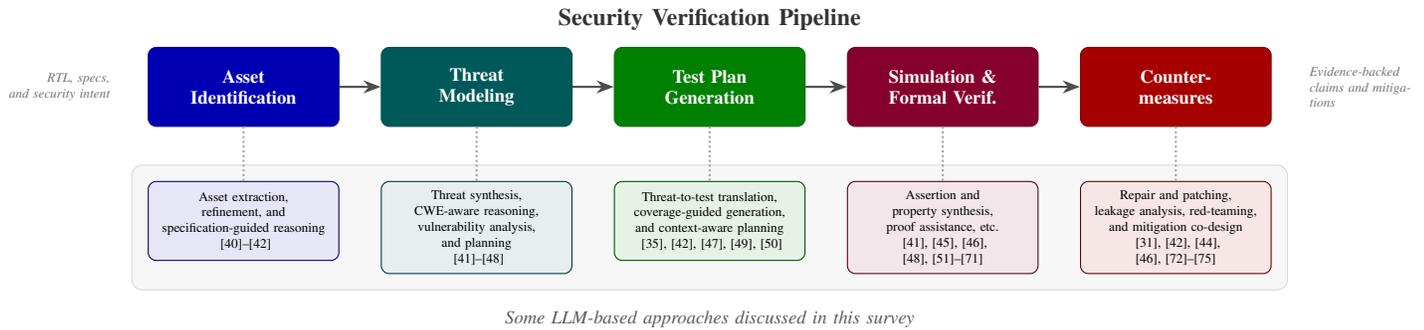

This paper surveys AI/LLM-based security verification along the key stages commonly followed by verification engineers: asset identification, threat modeling, security test-plan generation, construction of simulation and formal verification artifacts, and reasoning about potential countermeasures. We keep AI accelerators as the motivating domain throughout and use the open-source NVIDIA Deep Learning Accelerator (NVDLA) as a case-study platform, focusing on the CSB master module~\cite{nvdla_site}. Figure~\ref{fig:pipeline} illustrates the end-to-end verification pipeline that organizes both the survey and the case study.

\section{Survey of AI/LLM-Based Security Verification}
Recent research shows that AI/LLM-based techniques are increasingly applied across the hardware security verification workflow, from high-level threat reasoning to low-level assertion generation and debugging. Open hardware platforms and benchmark suites are also emerging to support evaluation of these approaches~\cite{nvdla_site,nvdla_hw_repo,gemmini_repo,opentitan_site,hardsecbench2025,fu2025hwfixbench,li2025fixbenchrtl}.

\subsection{Asset Identification}
Asset identification is the starting point of security verification~\cite{b4} because it determines what must be protected before threats, properties, and tests can be defined. Prior hardware-security research consistently places asset discovery at the front of the verification process, especially for confidentiality, integrity, and availability analysis, and shows that delayed identification often shifts security fixes to the far more costly post-silicon stage~\cite{7477271,7987680,8000621,mitra2010post}. Similar importance is reflected in broader cybersecurity and risk-analysis literature, where asset identification is treated as a prerequisite for systematic threat and vulnerability assessment~\cite{Yunizal2022AssetIdentification,6816371}.

Despite its importance, hardware asset identification has traditionally remained manual or dependent on predefined asset sets. Many prior works either assume assets are already known, begin from manually selected security-critical structures, or derive secondary assets only after being seeded with a small set of primary assets~\cite{meza2023informationflowcoveragemetrics,7858392,9441039,8242062,lasp2024}. Standards-oriented efforts such as SA-EDI and IEEE P3164 provide a principled vocabulary for conceptual and structural assets, but still leave substantial manual effort to the engineer~\cite{b4,b5}. More recent methods move toward partial automation through threat-informed or specification-guided reasoning, yet they remain sensitive to incomplete context and hand-crafted assumptions~\cite{11133104,10497111,11014404,lashed2025}.

The challenge becomes even greater in sophisticated systems such as AI accelerators, where assets extend beyond registers and control state to datasets, model parameters, intermediate activations, memory-resident data, and execution pathways that may leak or be manipulated through physical or logical attacks~\cite{10915219,11269794,10406065,10567471}. This growing complexity has motivated LLM-assisted asset-identification frameworks, with LAsset emerging as a representative effort to automate asset extraction and refinement across SoC and IP designs~\cite{hasan2026lassetllmassistedsecurityasset}.
\vspace{-2mm}

\subsection{Threat Modeling}
Threat modeling translates an asset inventory into adversarial scenarios that guide executable verification objectives. In hardware systems, this process is difficult because modern SoCs combine heterogeneous IPs, complex interconnects, and multiple privilege boundaries, creating a large and difficult-to-analyze attack surface. To structure this reasoning, researchers often use vulnerability taxonomies such as the Common Weakness Enumeration (CWE), which categorize weaknesses that may lead to exploitable vulnerabilities~\cite{CWE_mitre}. Mapping assets or design components to CWE categories helps engineers reason about likely attack vectors and their impact on confidentiality, integrity, and availability.

Earlier efforts explored automated vulnerability detection using CWE knowledge bases. Ahmad \textit{et al.} proposed a static-analysis framework~\cite{ahmad2022don} that scans Verilog RTL to detect potential CWE instances using rule-based scanners derived from abstract syntax trees. However, such approaches depend heavily on handcrafted rules and expert-defined patterns, which limit scalability.

Recent advances in LLMs have opened new opportunities for automating hardware security analysis~\cite{paria2023divas, lashed2025, 11022932, svllm2025}. For example, \textit{DIVAS}~\cite{paria2023divas} analyzes SoC specifications to identify vulnerabilities, map them to CWE entries, and generate security assertions, while \textit{LASHED}~\cite{lashed2025} combines static analysis with LLM reasoning for RTL vulnerability detection. More broadly, \textit{ThreatLens}~\cite{11022932} and \textit{SV-LLM}~\cite{svllm2025} extend this direction to agentic threat modeling, test-plan generation, and security property generation. Overall, these works show the growing role of LLM-driven automation in scalable hardware security verification, while challenges remain in accurate CWE mapping, large design reasoning, and reliable security analysis.
\vspace{-5mm}

\subsection{Test Plan Generation}

Traditional security test planning converts assets, threat hypotheses, and security objectives into executable stimuli, coverage targets, and measurement criteria. In established verification flows, the backbone remains SystemVerilog and the Universal Verification Methodology (UVM), while Portable Stimulus enables reuse of test intent across simulation and other execution platforms. Additional techniques such as coverage-directed exploration, test vector leakage assessment (TVLA), and learned automatic test pattern generation (ATPG) further extend these flows toward coverage closure, leakage evaluation, and activation of rare security conditions \cite{ieee1800_2017,accellera_uvm,accellera_pss,laeufer2018,goodwill2011,nn_atpg2024}.

AI/LLM-based planning approaches aim to bridge the gap between manual textual design intent and executable security tests. For example, ThreatLens and SV-LLM explicitly link threat reasoning to security test plan generation, while LLM4DV closes the loop by using coverage feedback to generate new stimuli for uncovered behaviors \cite{11022932,svllm2025,zhang2023llm4dv}. ChipNeMo~\cite{chipnemo2023} and GPT4AIGChip~\cite{10323953} further suggest that domain-adapted or accelerator-aware LLM workflows can improve planning quality when test intent is inferred primarily from architecture documents rather than mature security specifications. Taken together, these works make test-plan generation one of the most practical near-term uses of AI/LLM methods because the outputs remain structured, reviewable, and closely tied to the threat model.
\subsection{Simulation-Driven and Formal Verification}
Simulation and formal verification remain at the core of trustworthy security verification. Conventional practice relies on SystemVerilog simulation infrastructure~\cite{clarke1999}, assertion-based verification~\cite{tiwari2009}, model checking~\cite{hu2011}, information-flow tracking~\cite{ieee1800_2017}, and coverage-guided exploration~\cite{laeufer2018} to determine whether suspected weaknesses can actually manifest in design. In AI/LLM-based verification, generated properties, monitors, and hypotheses are useful only when they compile, execute correctly, and meaningfully constrain the design.

Recent work increasingly explores LLM-assisted assertion generation. Early frameworks such as AutoSVA~\cite{orenesvera2021autosva} and its later extension~\cite{orenesvera2023rtl2sva} automate formal property generation from RTL and verification feedback. Methods such as LISA~\cite{paul2025lisa} and STELLAR~\cite{rajabi2025stellar} improve assertion quality through retrieval-augmented reasoning, structural grounding, and guided prompting, while AssertionLLM~\cite{pulavarthi2025assertionllm} highlights the limitations of off-the-shelf models and proposes a fine-tuned alternative. 

Beyond assertion synthesis, AI/LLM methods are increasingly integrated into broader verification workflows. DIVAS~\cite{paria2023divas}, LASA~\cite{ankireddy2025lasa}, and LLM-IFT~\cite{11022949} move toward end-to-end security analysis by generating properties and structured information-flow reasoning from RTL, specifications, and design documents. Bug-detection and debugging frameworks such as BugWhisperer~\cite{11022958}, LASHED~\cite{lashed2025}, and MEIC~\cite{meic2024} further incorporate iterative tool-in-the-loop refinement to localize, explain, and repair RTL vulnerabilities.

\subsection{Countermeasures}
Countermeasures form the defensive endpoint of the verification pipeline, because discovered vulnerabilities must ultimately be prevented, detected, or tolerated in deployment. Classical hardware-security defenses include masking and hiding for side-channel resistance~\cite{ishai2003}, fault detection and redundancy against active fault injection~\cite{barenghi2012} and logic locking for IP protection~\cite{rajendran2012}.

Recent research suggests that AI/LLMs can also support mitigation generation by helping engineers identify RTL vulnerabilities, reason about their root causes, and explore design-level repair options. Ahmad \textit{et al.} show that prompted LLM ensembles can generate candidate fixes that outperform prior automated hardware-repair baselines on a set of security bugs~\cite{10462177 }. At a broader level, frameworks such as LLM-Sec~\cite{11125787} illustrate how LLMs can assist with vulnerability analysis, policy reasoning, and remediation planning. Overall, the emerging direction is a closed-loop hardware-security workflow in which AI not only helps expose weaknesses, but also supports continuous security-aware design refinement.

\section{Research Questions}
Recent benchmark efforts and open hardware platforms have begun to establish an initial foundation for evaluating AI/LLM-based security verification. At the same time, they expose unresolved challenges in accelerator-aware abstractions, reliable ground truth, and rigorous artifact validation~\cite{hardsecbench2025,fu2025hwfixbench,li2025fixbenchrtl,nvdla_hw_repo,gemmini_repo,opentitan_site}. Having viewed collectively, the survey literature points not only to emerging capabilities, but also to a set of broader questions that remain open for the community.
\begin{enumerate}[leftmargin=*]
\item \textbf{RQ1.} How can AI/LLM-based methods automate the end-to-end security verification workflow for hardware and AI accelerators?
\item \textbf{RQ2.} How can simulation, formal methods, and structured security knowledge ground AI/LLM-generated verification artifacts and reduce hallucination?
\item \textbf{RQ3.} What benchmarks, evaluation criteria, and accelerator-specific abstractions are required to make AI/LLM-based security verification practical and trustworthy at scale?
\end{enumerate}

\section{Case Study of NVDLA}

To demonstrate the practical application of the research questions, we present a case study on the open-source NVDLA accelerator, focusing on the \texttt{NV\_NVDLA\_csb\_master} IP. As the configuration ingress point for CSB requests before decoding and forwarding, this module provides a natural pre-silicon boundary for access-control analysis, where unauthorized requests can be examined before downstream effects occur. This case study does not claim that every NVDLA deployment is exploitable; rather, it evaluates whether the CSB master, viewed in isolation, lacks local privilege enforcement, and therefore relies on trusted integration or upstream filtering. The analysis is thus intentionally scoped to this IP boundary. The tables and artifacts reported here are condensed outputs of an automated agentic workflow. All RTL line references correspond to the public \texttt{nvdlav1} snapshot of the NVDLA CSB master\footnote{\url{https://github.com/nvdla/hw/tree/nvdlav1/vmod/nvdla/csb_master}}.
\vspace{-2mm}
\subsection{Asset Identification}
Given the CSB master boundary defined above, the next step is to identify the assets that govern request acceptance and forwarding. For this case study, the most relevant assets are the CSB handshake, the request payload fields, the internal address-decode logic, and the decoded DMA-facing interfaces. Together, these determine whether an unauthorized requester can influence security-critical behavior. Table~\ref{tab:assets} summarizes the assets considered in the threat analysis. Although the exercised assets are primarily control and integrity-oriented, unauthorized control of DMA-facing paths can also create confidentiality risk by enabling exposure or redirection of memory-resident model parameters, activations, or outputs. Additionally, the concentration of CWE-1220 mappings across the request and decode boundary motivates the threat framing summarized next in Table~\ref{tab:threat-model}.

\vspace{-5mm}

\begin{table}[!ht]
\centering
\caption{LLM-generated security assets for the CSB master}
\label{tab:assets}
\footnotesize
\renewcommand{\arraystretch}{1.1}
\begin{tabularx}{\columnwidth}{@{}>{\raggedright\arraybackslash}p{0.34\columnwidth}>{\centering\arraybackslash}p{0.08\columnwidth}X@{}}
\hline
\textbf{Asset} & \textbf{Obj.} & \textbf{Mapped CWEs} \\
\hline
CSB valid/ready handshake & A & CWE-1318, CWE-440, CWE-1429 \\
CSB write data & I & CWE-1262, \textbf{CWE-1220}, CWE-1318 \\
CSB write control & I & CWE-1262, \textbf{CWE-1220}, CWE-1318 \\
Internal address select logic & I & CWE-1276, CWE-1262, \textbf{CWE-1220} \\
CSB-to-CVIF interface & I, A & CWE-1262, \textbf{CWE-1220}, CWE-1318 \\
\hline
\multicolumn{3}{@{}p{\columnwidth}@{}}{\scriptsize Obj.: I = Integrity, A = Availability.}
\vspace{-3mm}
\end{tabularx}
\vspace{-3mm}

\end{table}

\subsection{Threat Modeling}
At this boundary, the central question is whether any requester reaching the CSB master can access decoded sub-unit paths without a privilege context. This is a meaningful module-level concern because \texttt{core\_req\_prdy} is permanently asserted (line~468), while no local authorization check is applied to source privilege information at this boundary. Table~\ref{tab:threat-model} summarizes this condition in terms of the request format, threat agent, attack vector, STRIDE mapping, and scope. This threat framing directly determines the directed tests summarized in Table~\ref{tab:test-plan}.
\vspace{-3mm}
\begin{table}[!ht]
\centering
\caption{LLM-generated threat model for CWE-1220.}
\label{tab:threat-model}
\footnotesize
\renewcommand{\arraystretch}{1.1}
\begin{tabular}{@{}p{0.16\columnwidth}p{0.78\columnwidth}@{}}
\hline
\textbf{Field} & \textbf{Detail} \\
\hline
CWE ID & CWE-1220: Insufficient Granularity of Access Control \\
Root Cause & \texttt{core\_req\_prdy = 1'b1}; no privilege check in decode logic \\
Format & 50-bit packet: \{nposted, write, addr[15:0], wdata[31:0]\}; no security field \\
Threat & Untrusted requester can issue accepted CSB read/write requests \\
Scenario & Discover address map, issue unprivileged writes, observe forwarding \\
STRIDE & Tampering, Information Disclosure, Elevation of Privilege \\
Scope & Module boundary only; downstream effects not claimed \\
\hline
\end{tabular}
\end{table}
\vspace{-3mm}
\subsection{Test Plan Generation}
The threat model leads to a small set of directed tests centered on a single question: does the CSB master accept and forward the exercised requests? Table~\ref{tab:test-plan} summarizes the executed plan, focusing on DMA-facing targets and the broader decoded address space. All addresses follow the CSB master's word-address convention, and each test maps directly to a SystemVerilog testbench task.

\vspace{-5mm}
\begin{table}[!ht]
\centering
\caption{LLM-generated test plan for CWE-1220.}
\label{tab:test-plan}
\footnotesize
\renewcommand{\arraystretch}{1.1}
\begin{tabular}{@{}p{0.02\columnwidth}p{0.26\columnwidth}p{0.62\columnwidth}@{}}
\hline
\textbf{\#} & \textbf{Target / Op} & \textbf{Check} \\
\hline
1 & MCIF write & Forwarding of \texttt{0x0800}, \texttt{0xDEAD\_BEEF} to MCIF \\
2 & BDMA/CVIF/CDMA write & Forwarding to decoded DMA-facing targets without privilege check \\
3 & MCIF read & Request accepted at the same boundary \\
4 & Full scan (18) & Acceptance across 17 mapped targets + 1 unmapped \\
5 & Rapid-fire ($\times$8) & No throttling or back-pressure on MCIF writes \\
\hline
\end{tabular}
\end{table}

\balance

\begin{lstlisting} [style=artifact,language=Verilog,caption={Representative LLM-generated stimuli corresponding to the LLM-generated plan.},label={lst:testbench-stimuli}]
// Test 1: Unprivileged MCIF write
task automatic test1_unpriv_write_mcif();
  u_drv.drive_txn(
    .addr(ADDR_MCIF),           // 16'h0800
    .wdata(32'hDEAD_BEEF),
    .write(1'b1), .nposted(1'b1),
    .description("UNPRIV WRITE forwarded to MCIF"));
endtask

// Test 4: Full 18-target address scan
task automatic test4_full_address_scan();
  localparam logic [15:0] test_addrs [18] =
    '{ADDR_GLB,  ADDR_GEC,  ADDR_MCIF, ADDR_CVIF, ADDR_BDMA, ADDR_CDMA, ADDR_CSC,  ADDR_CMAC_A, ADDR_CMAC_B, ADDR_CACC, ADDR_SDP_RDMA, ADDR_SDP, ADDR_PDP_RDMA, ADDR_PDP, ADDR_CDP_RDMA, ADDR_CDP, ADDR_RBK, ADDR_UNMAPPED};
  for (int i = 0; i < 18; i++)
    u_drv.drive_txn(.addr(test_addrs[i]), ...);
endtask
\end{lstlisting}
Artifact~\ref{lst:testbench-stimuli} shows representative task code for the direct MCIF write and the broader address scan. In the executed bench, these tasks drive requests through an async FIFO modeling the Falcon--core clock-domain crossing, while monitors track ingress acceptance and decoded forwarding.
\subsection{Simulation Driven Verification}
The test plan is then evaluated through simulation. The exploit-oriented bench drives the planned transactions at the CSB ingress and records handshake activity and decoded forwarding using a driver, request monitor, sub-unit monitor, and scoreboard. Of the 31 exercised transactions, 30 were observed at decoded sub-unit ports and flagged as security violations (29~writes and 1~read). The remaining transaction targeted the unmapped address (\texttt{0x7C00}); it was accepted at ingress, but routed to an internal dummy client that is not externally monitored. Across all valid requests, the ready signal never deasserted, indicating zero back-pressure in the exercised cases. These results support a module-scoped conclusion: at this boundary, the CSB master accepts and forwards the tested requests without a local privilege check; whether this becomes a system-level vulnerability or not depends on SoC integration and reachability of the CSB path.

\subsection{Countermeasures}
This behavior points to mitigation at the request-admission boundary. Because the issue arises before decoded forwarding begins, the most direct response is to constrain request admission in the CSB master rather than retrofit each downstream target independently. Table~\ref{tab:countermeasure} summarizes the relevant control point and corresponding design options.
\vspace{-5mm}
\begin{table}[!ht]
\centering
\caption{RTL control point and suggested countermeasures}
\label{tab:countermeasure}
\footnotesize
\renewcommand{\arraystretch}{1.1}
\begin{tabular}{@{}p{0.24\columnwidth}p{0.66\columnwidth}@{}}
\hline
\textbf{Item} & \textbf{Detail} \\
\hline
Control Point & \texttt{core\_req\_prdy} is permanently high (line~468), so all requests are accepted and may be forwarded downstream. \\
\hline
\multicolumn{2}{@{}l}{\textbf{Recommendations}} \\
\hline
R1 & Enforce interconnect-level access control before this module. \\
R2 & Add privilege or source-ID metadata for local authorization. \\
R3 & Gate \texttt{core\_req\_prdy} on an authorization check. \\
\hline
\end{tabular}
\end{table}

These recommendations matter when the CSB path may be reachable by untrusted requesters in a larger SoC integration.

\section{Conclusion}
In this paper, we've surveyed AI/LLM-based methods for hardware security verification, with AI accelerators as the motivating domain. Although the literature increasingly spans the workflow from asset identification and threat modeling to test generation, verification artifacts, and countermeasure support, practical adoption still depends on grounding through executable evidence, formal methods, benchmark quality, and accelerator-aware evaluation. The NVDLA case study demonstrates that an automated agentic workflow can support a scoped, simulation-grounded RTL security analysis and produce localized mitigation guidance.


\bibliographystyle{IEEEtran}
\bibliography{bib_file}

\end{document}